\documentclass[
aps,%
12pt,%
final,%
notitlepage,%
oneside,%
onecolumn,%
nobibnotes,%
nofootinbib,%
superscriptaddress,%
noshowpacs,%
centertags]%
{revtex4}

\begin{document}

\title{Optimal Filtration and a Pulsar Time Scale}

\author{\firstname{A.E.}~\surname{Rodin}}

\email{rodin@prao.ru} \affiliation{ Pushchino Radio Astronomy
  Observatory, Astro Space Center, Lebedev Physical Institute}
\author{\firstname{Ding}~\surname{Chen}}

\email{ding@cssar.ac.cn}
\affiliation{
Center for Space Science and Applied Research, Chinese Academy of Sciences
}

\begin{abstract}
An algorithm is proposed for constructing a group (ensemble) pulsar
time based on the application of optimal Wiener filters. This algorithm
makes it possible to separate the contributions of variations of the
atomic time scale and of the pulsar rotation to barycentric residual
deviations of the pulse arrival times. The method is applied to
observations of the pulsars PSR B1855+09 and PSR B1937+21, and is used
to obtain corrections to UTC relative to the group pulsar time
PT$_{\rm ens}$.  Direct comparison of the terrestial time TT(BIPM06)
and the group pulsar time PT$_{\rm ens}$ shows that they disagree by
no more than $0.4\pm0.17\; \mu$s. Based on the fractional instability of
the time difference TT(BIPM06) -- PT$_{\rm ens}$, a new limit for the
energy density of the gravitational-wave background is established at
the level $\Omega_g {h}^2\sim 10^{-9}$.
\end{abstract}

\maketitle

The discovery of pulsars in 1967 \cite{Hewish68} and of millisecond
pulsars in 1982 \cite{Backer82}, as well as subsequent observations,
have clearly shown that the rotational stability of pulsars is such
that pulsars can be used as astronomical clocks. Currently, the
accuracy with which pulse arrival times (PATs) can be measured is at
the level of several microseconds for most pulsars, and reaches the
submicrosecond level for some pulsars. If we compare this accuracy to
the interval covered by observations, we obtain the relative accuracy
$10^{-15}$, which is essentially comparable to the fractional
instability of the best atomic frequency standards.

A number of studies have been concerned with the stability of pulsar
rotation and its application to time systems. Writing about the
principles for establishing Terrestrial Time (TT), Guinot
\cite{guinot88} concludes that Terrestrial Time as realized by the
Bureau International de Poids et Mesures [TT(BIPMXX), where XX denotes
  the year] is the most expedient time system to use for pulsar
chronometry. The principles of pulsar time and the definition of the
pulsar second are given in \cite{ilyasov89}. It is shown in
\cite{guinot91} that it is not possible to use the rotation of pulsars
to improve the definition of the time unit. The studies
\cite{ilyasov96,kopeikin97,rodin97,ilyasov98} introduce a definition
for a time based on the motion of a pulsar in a binary system --
Binary Pulsar Time, or BPT -- together with theoretical expressions
for the variations in the rotational and orbital parameters as
functions of the duration of the observing interval. The main
conclusion of these studies is that BPT is less stable than the
ordinary pulsar time PT over short intervals, but the fractional
instability of BPT can reach $10^{-14}$ over long intervals ($10^2\div
10^3$ yr). Petit and Tavella \cite{petit96} present an algorithm for
determining a group pulsar time based on weighted averaging, which is
compared to an algorithm based on optimal filtration; they also
consider some ideas connected with the stability of BPT. Foster and
Backer \cite{foster90} present a polynomial approach to describing
clock and ephemerides variations and the influence of gravitational
waves passing through the solar system on pulsar chronometry.

Here, we propose a method for forming a group pulsar time (PT) based
on the application of optimal Wiener filters \cite{rodin06,rodin08}. Section 2
discusses the pulsar-chronometry (timing) algorithm from the point of
view of time scales. Section 3 describes the principles for
constructing optimal filters and the formation of a group pulsar
time. Section 4 contains the results of computer simulations of
detecting a signal against the background of correlated noise using
optimal filtration. Section 5 discusses the results of applying Wiener
filters to observations of the pulsars PSR B1855+09 and PSR B1937+21.

\section{PULSAR CHRONOMETRY ALGORITHM}

An observer located on the Earth, which is rotating about its axis and
revolving around the Sun, receives a signal from a pulsar using a
radio telescope over an accumulation time that is sufficient to obtain a
specified signal-to-noise ratio. The pulse arrival time (PAT) is
measured in a local time scale via crosscorrelation with a standard
profile for the pulsar pulse.  The resulting PATs $\tau_N$ can be translated
into UTC, TAI, and TT using the relations \cite{oleg90}
\begin{equation}
{\rm UTC}=\tau_N+\Delta\tau,\;{\rm TAI}={\rm UTC}+k,\; {\rm TT}={\rm
TAI}+32.184\,{\rm s},
\end{equation}
where  $\Delta\tau$ is the local time correction, $k$ is a whole
number of seconds introduced to take into account
the variation in the length of the day, and the quantity
32.184 s is added to remove a historical jump between
atomic and ephemerides time. Since the duration of
the second in TT depends on the position and velocity
of the Earth in its orbit, an additional transformation
from TT to Barycentric Time TB is required, using the
algorithm described in \cite{fairhead90}.

The observations that have been transformed into
the uniform Barycentric Time system TB must then
be reduced to the barycenter of the solar system, in
order to exclude variations in the PATs due to the
motion of the observer about the barycenter \cite{oleg90}
\begin{equation}
T=t-t_0+\Delta_R(\alpha,\delta,\mu_\alpha,\mu_\delta,\pi)+\Delta_{\rm
orb}-DM/f^2+\Delta_{\rm rel},
\end{equation}
where $t_0$ is the initial epoch;  $t$ the topocentric PAT in TB; $T$
the PAT at the barycenter of the solar system;
$\Delta_R(\alpha,\delta,\mu_\alpha,\mu_\delta,\pi)$ the Remer
correction along the Earth's orbit;
$\alpha,\delta,\mu_\alpha,\mu_\delta$ and $\pi$ the right ascension,
declination, proper motion in right ascension and declination, and
parallax of the pulsar, respectively; $\Delta_{\rm orb}$ the Remer
correction along the orbit of the pulsar, when it is located in a
multiple system; $DM/f^2$ the delay for the signal propagation in the
interstellar and interplanetary media at frequency $f$ taking into
account the Doppler shift; and $\Delta_{\rm rel}$ the relativistic
correction for the delay of the signal propagation in the
gravitational field of the solar system.

The PATs at the solar-system barycenter are then
used to calculate the rotational phase (number of
rotations) of the pulsar:
\begin{equation}\label{phase}
N(T)=N_0+\nu T +\frac12\dot\nu{T}^2+\varepsilon(T),
\end{equation}
where  $N_0$ is the initial phase at time $T=0$, $\nu$ and $\dot\nu$
are the rotational frequency of the pulsar and its derivataive at
$T=0$, and $\varepsilon(T)$ represents variations in the rotational
phase (timing noise). The procedure for determining the parameters
includes refining $\alpha,\delta,\mu_\alpha,\mu_\delta$ and $\pi$, via
a least-squares fit minimizing a weighted sum of the square differences
between $N(T)$ and the nearest integer. The residual deviations in the
rotational phase are usually expressed in units of the time $\delta
t=\delta N/\nu$. Here, we consider variations in the intrinsic
rotation of the pulsar and variations due to irregularity of the
reference time scale $\Delta t_{clock}(T)$.

When intercomparing different realizations of atomic time
\cite{guinot88}, flicker noise in the frequency dominate on intervals
of several months, while random walk noise in the frequency dominates
on intervals of several years. Thus, clock variations have a power
spectrum of the form $1/\omega^{n}$ in the frequency domain, and are
expressed in the time domain by the polynomial
\begin{equation}
\Delta {t}_{\rm clock}({\cal T})=c_0 + c{\cal T}+\frac 12\dot c{\cal
T}^2+ \frac 16\ddot
 c{\cal T}^3+\ldots.
\end{equation}
It is clear that the appearance of the quantity $\Delta t_{\rm clock}$
in Eq. (\ref{phase}) leads to a redetermination of the rotational
parameters of the pulsar:
\begin{equation}
N({\cal T})={N}_0'+(1+c)f {\cal T}+\frac 12(f\dot c+(1+c)^2\dot
f){\cal T}^2,
\end{equation}
where $\cal T$ is ideal barycentric time and $f$ and $\dot f$ are the
rotational frequency of the pulsar and its derivative undistorted by
the clock parameters. For this reason, we must use the PAT values
expressed in TT(BIPMXX), as best determined at the current time
\cite{guinot91}.

\section{OPTIMAL FILTRATION}

Let us consider $n$ uniform measurements of a random quantity (the
residual PAT deviations)  ${\bf r} = (r_1,r_2,\ldots,r_n)$. The
quantity ${\bf r}$ is the sum of two uncorrelated quantities, ${\bf
  r}={\bf s}+\varepsilon$, where ${\bf s}$ is a random signal to be
estimated and associated with the clock contribution, and
$\varepsilon$ is the contribution of the rotational phase of the
pulsar, which we treat here like noise. We are interested in
estimating the signal ${\bf s}$ against the background of the additive
noise $\varepsilon$ using the Wiener filtration method.

Wiener filtration consists of estimating the signal ${\bf s}$ given
the measurements ${\bf r}$ and covariances (\ref{cov})
\cite{gubanov97}.  We must reconstruct the random signal with
insufficient a priori information, since the covariance matrix of the
signal is not known a priori, and is estimated from the observational
data themselves via a crosscorrelation of all the data, assuming that
the variations in the clicks (the estimated signal) and in the pulsar
rotational phase (additive noise) are uncorrelated.

The optimal Wiener estimate of the signal ${\bf s}$ is given by \cite{gubanov97}:

\begin{equation}\label{signal}
{\bf\hat s}={\bf Q}_{sr}^{}{\bf Q}_{rr}^{-1}{\bf r}
={\bf Q}_{ss}^{}{\bf Q}_{rr}^{-1}{\bf r}
={\bf Q}_{ss}({\bf Q}_{ss}+{\bf Q}_{\varepsilon \varepsilon})^{-1}{\bf
r},
\end{equation}
where the covariance matrices ${\bf Q}_{rr}$, ${\bf Q}_{sr}$, and
${\bf Q}_{ss}$ are formed as arrays of the corresponding covariance
functions. The covariance functions for  ${\bf r}$, ${\bf s}$ and
$\varepsilon$ are written

\begin{equation}\label{cov}
\begin{array}{l}
{\rm cov}(r,r)=\left<r_i,r_j\right>= \left<s_i,s_j\right>+
\left<\varepsilon_i, \varepsilon_j\right>,\\
{\rm cov}(s,s)=\left<s_i,s_j\right>,\\
{\rm cov}(s,r)=\left<s_i,r_j\right>=\left<r_i,s_j\right>
= \left<s_i,s_j\right>,\\
{\rm cov}(\varepsilon,\varepsilon)=\left<\varepsilon_i,
\varepsilon_j\right>.
\end{array}\;(i,j=1,2,\ldots,n).
\end{equation}
The angular brackets, $\left< \right>$, denote ensemble averaging.  We
assume that the processes  $s$ and $\varepsilon$ are stationary in a
weak sense; i.e., only the first and second moments are stationary,
since fitting a quadratic polynomial to the rotational phase excludes
the nonstationary part of the random process \cite{kopeikin99}. Thus,
the covariance functions depend only on the time difference $t_i-t_j$. 

Since the pulsar-chronometry algorithm assumes that the PATs are
defined relative to a reference time, distinguishing the covariances
$\left<s_i,s_j\right>$ and $\left<\varepsilon_i,\varepsilon_j\right>$
requires observations of at least two pulsars using a single time
system. In this case, combining the pulsar PATs and assuming that the
cross-covariance $\left<{}^2\varepsilon_i,{}^1\varepsilon_j\right> =
\left<{}^1\varepsilon_i,{}^2\varepsilon_j\right>=0$, we can estimate
\begin{equation}
\left<s_i,s_j\right> =
\left(\left<{}^1r_i+{}^2r_i,{}^1r_j+{}^2r_j\right> -
\left<{}^1r_i-{}^2r_i,{}^1r_j-{}^2r_j\right>\right)/4
\end{equation}
or
\begin{equation}
\left<s_i,s_j\right> = \left<{}^1r_i,{}^2r_j\right>.
\end{equation}
If M pulsars are observed to construct a group pulsar time, we will
have $M(M-1)/2$ estimates of the covariance matrices
$\left<s_i,s_j\right> = \left<{}^k r_i,{}^l r_j\right>,\;
(k,l=1,2,\ldots,M)$

The matrix  ${\bf Q}_{rr}^{-1}$ in (\ref{signal}) is a whitening
filter. The matrix ${\bf Q}_{ss}$ forms a signal from the whitened
data.

The averaged signal (group pulsar time) is written
\begin{equation}\label{ensemble}
{\bf\hat s_{\rm ens}} = \frac{2}{M(M-1)}
\sum_{m=1}^{\frac{M(M-1)}2}{}_m{\bf Q}_{ss}\cdot
\sum_{i=1}^{M}{}{^i}w\;{}^i{\bf Q}_{rr}^{-1} \cdot{}^i{\bf r},
\end{equation}
where $^i w$ is the relative weight of the $i$th pulsar, $^iw=
\kappa/\sigma_i^2$, $\sigma_i$ is the rms deviation of the whitened
data ${}^i{\bf Q}_{rr}^{-1} \cdot{}^i{\bf r}$, and the constant
$\kappa$ serves to ensure that $\sum_i\, {^iw}=1$. The first factor in
(\ref{ensemble}) is the average cross-covariance function, and the
second factor is the weighted sum of the whitened data.

The following algorithm was used to calculate the auto- and
cross-covariances. The observational data $^kr_t$ were subjected to a
rapid Fourier transform,
\begin{equation}
{}^kx(\omega)=\frac{1}{\sqrt{n}}\sum_{t=1}^n {}^kr_t h_t
e^{i\omega t},\;(k=1,2,\ldots,M),
\end{equation}
where the weights $h_t$, which are bell-like functions, were used to
reduce leakage through the sidelobes \cite{percival91}. They can be
calculated to very good accuracy using the formula \cite{percival91}
\begin{equation}
h_t=C_0'\frac{I_0\left(\pi W(n-1)\sqrt{\left[1-(\frac{2t-1}{n}-1)^2
\right]} \right)}{I_0(\pi W(n-1))},
\end{equation}
where $C_0'$ are scaling factors used to ensure that $\sum h_t^2=1$,
$I_0$,  is a modified zeroth-order Bessel function of the first kind,
and the parameter $W$ serves to regularize the sidelobe level in the
spectrum, and is usually taken to be in the range $W=1 \div 4$. We
used the value $W=1$.

The power spectrum ($k=l$) and cross-spectrum ($k\neq l$) were
calculated using the formula
\begin{equation}\label{spxy}
{}^{kl}X(\omega)=\frac{1}{2\pi}\left|^kx(\omega)^lx^{*}(\omega)
\right|,\end{equation}
where  $(\cdot)^*$ denotes complex conjugation.

Finally, the auto-  ($k=l$) and cross-  ($k\neq l$) covariance were
calculated using the formula
\begin{equation}
{\rm cov}(^kr,^lr) =\sum_{\omega=1}^n {^{kl}}X(\omega)
e^{-i\omega t},\;(k,l=1,2,\ldots,M).
\end{equation}

\section{COMPUTER MODELING}

We carried out numerical simulations to estimate the quality of the
signal reconstruction applying Wiener filtration and weighted
averaging \cite{petit96}. In constrast to \cite{rodin08}, where a
harmonic signal was taken as the initial signal, we estimated a time
series of the difference UTC--TT(BIPM06) in the interval MJD =
46399--48949, from which we subtracted the quadratic polynomial from
the least-squares fit.  We added white noise with zero mean and
various dispersions. For example, if we used 50 pulsars for the
modeling, the dispersion of the white noise $n_0$ was
$\sigma^2=1\div50$. The weights were taken to be inversely
proportional to the dispersion of the obtained noise series. The
quality of the reconstructed signal was calculated as the rms
deviation of the difference between the reconstructed and input
signals.  

Figures 1a and 1b show the quality of the signal
reconstruction using the weighted averaging (dashed curve) and Wiener
filtration (solid curve) for white noise, as functions of the number of
pulsars used and the length of the data series. As the length of the
data series and the number of pulsars increase, the quality of the
Wiener-filtration signal reconstruction becomes increasingly higher
compared to the weighted-average reconstruction as the rotational phase
variations increase.

\begin{figure}\label{fig1}
\setcaptionmargin{1mm}
\begin{minipage}{0.5\linewidth}
\vbox{\hbox{\hspace*{4.4cm}(a)}\includegraphics[width=8cm]{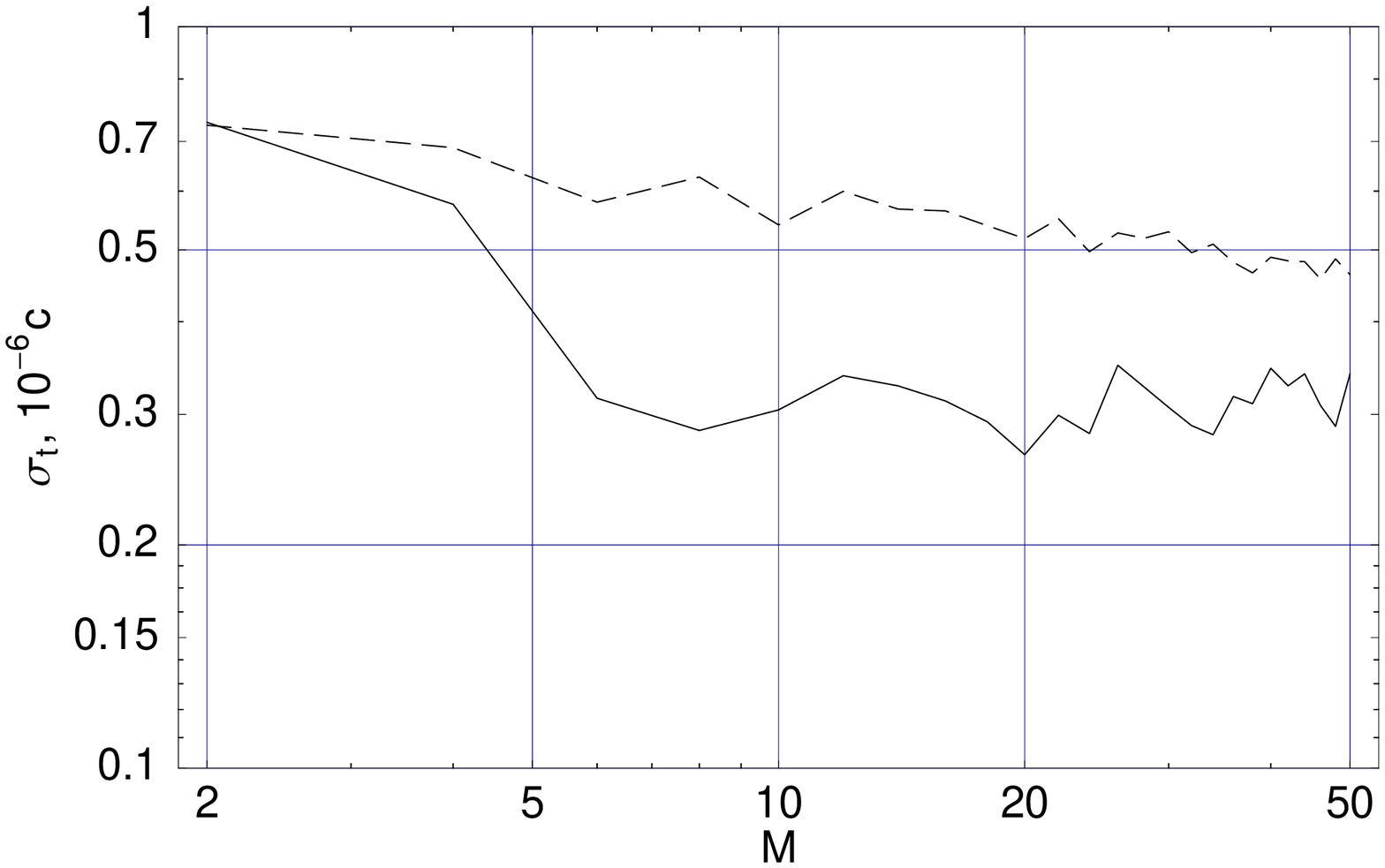}}
\end{minipage}\hfill
\begin{minipage}{0.5\linewidth}
\vbox{\hbox{\hspace*{4.4cm}(b)}\includegraphics[width=8cm]{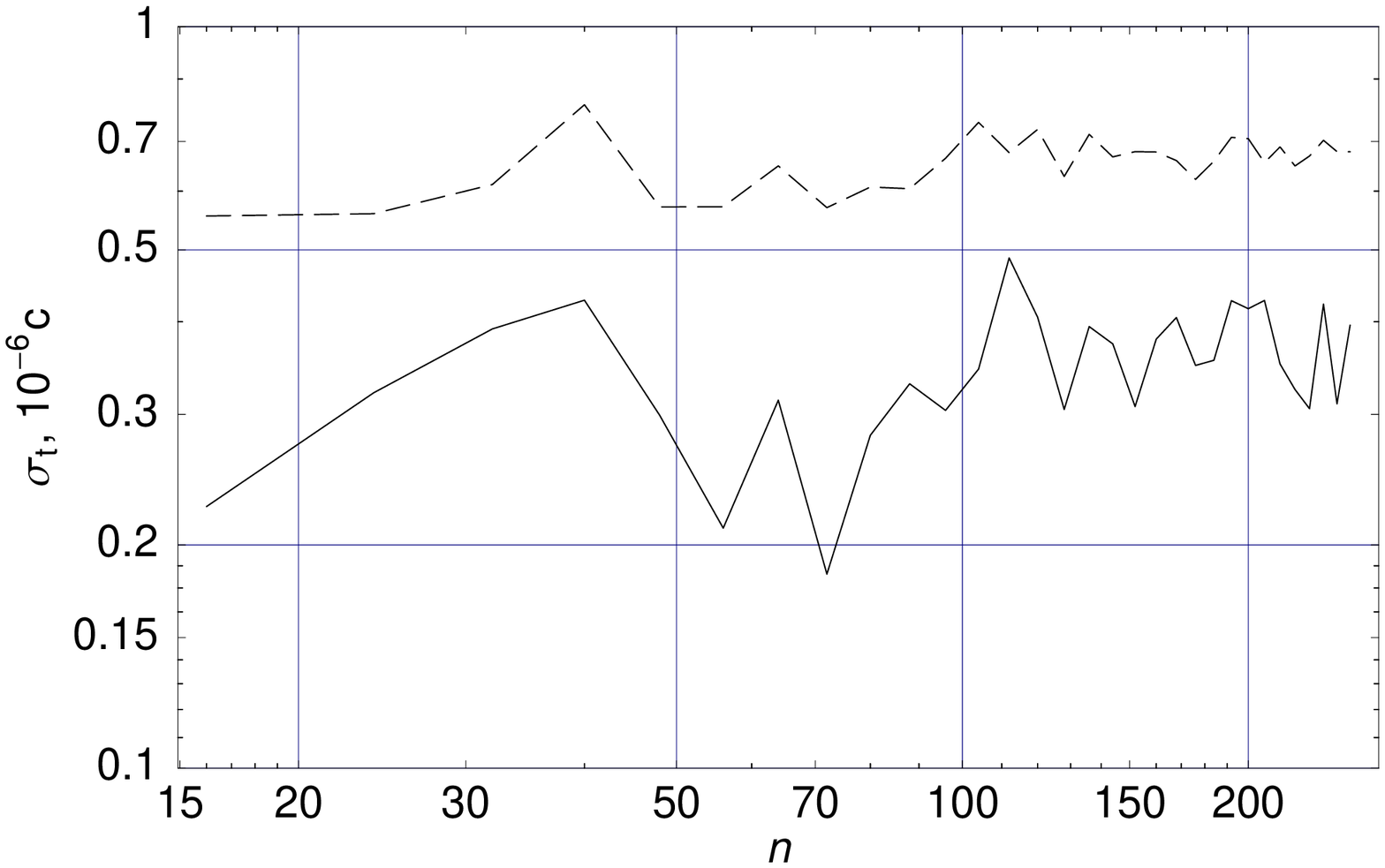}}
\end{minipage}
\captionstyle{normal}
\caption{The rms difference between the reconstructed and initial
  signals as a function of (a) the number of pulsars $M$ and (b) the
  length of the data series $n$, using weighted averaging (dashed) and
  Wiener filtration (solid).}
\end{figure}

\section{RESULTS}

The method described above was applied to observational data
(barycentric residual deviations of the PATs) for the pulsars PSR
B1855+09 and PSR B1937+21  \cite{kaspi94}. Since these data are not uniform, we
applied a spline interpolation to regularize the data, with the aim of
simplifying the subsequent reduction using matrix-algebra methods. We
chose a step of 10 days, which preserved the original quantity of
data. This transformation of the data distorts the high-frequency
component, but leaves the low-frequency component that is of interest
to us unchanged \cite{roberts87}.

We took the part of the data that was common to both pulsars (251
points for each), in the interval MJD = 46450--48950, for this
reduction. Since, generally speaking, the series of residual
deviations have different means and slopes after their lengths are
shortened, they were reduced by fitting a quadratic polynomial
(Fig. 2).

\begin{figure}\label{fig2}
\setcaptionmargin{5mm}
\onelinecaptionstrue
\includegraphics[width=12cm]{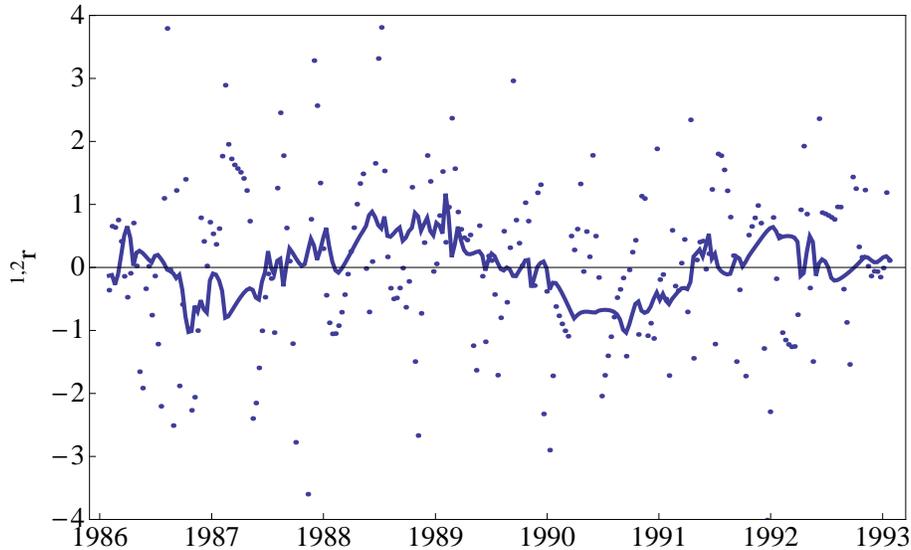} 
\captionstyle{normal}
\caption{Barycentric residual deviations ${}^{1,2}r$  (in $\mu$s) for
  PSR B1855+09 (points) and PSR B1937+21 (curve) after fitting a
  quadratic polynomial.}
\end{figure}

The data for PSR B1855+09 and PSR B1937+21 were obtained in UTC based
on the information from [19]. Consequently, the signals that were
distinguished from the observational data for PSR B1855+09 and PSR
B1937+21 were the corrections UTC--PT$_{1855}$ and UTC--PT$_{1937}$ . The combined
signal (group time) UTC--PTens is shown in Fig. 3 by the thin curve. It
shows behavior similar to that for UTC--TT(BIPM06), with the
correlation coefficient  $\rho=0.75\pm 0.04$. A direct comparision of
terrestrial time TT(BIPM06) and the group Pulsar Time PT$_{\rm ens}$ shows that
they disagree by no more than $0.4 \pm 0.17\; \mu$s.

We calculated the fractional instability $\sigma_z$ of the time
difference TT--PT$_{\rm ens}$, which was $\sigma_z=(0.5\pm 2)\cdot
10^{-15}$ on a time interval of seven years.
\begin{figure}\label{fig3}
\setcaptionmargin{5mm}
\onelinecaptionstrue
\includegraphics[width=12cm]{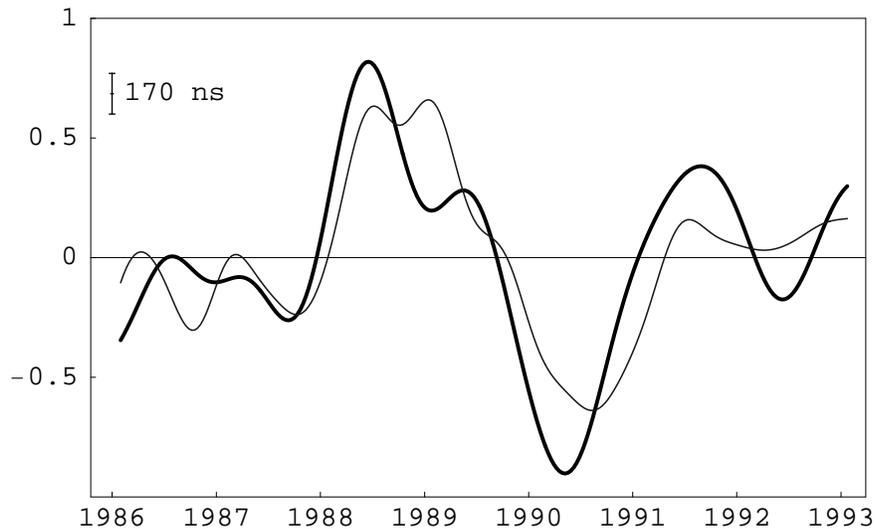} 
\captionstyle{normal}
\caption{UTC--TT(BIPM06) (in $\mu$s, bold curve) and UTC--PT$_{\rm
    ens}$ (thin curve).}
\end{figure}

\section{DISCUSSION}

The fractional instability of time systems is characterized by the
so-called Allan variance, which is numerically equal to the
second-order finite difference of the clock phase variations. A
chronometric analysis of observational data contains a determination
of the rotational parameters of the pulsar, including at least the
first frequency derivative; this corresponds to fitting a quadratic
polynomial in time to the rotational phase of the pulsar, which, in
turn, is equivalent to excluding the second-order derivative from the
PAT data. For this reason, the use of the classical Allan variance is
not expedient. Instead, another quantity was proposed as a means of
characterizing the fractional instability of the rotation of pulsars --
$\sigma_z$ \cite{taylor91}. A detailed algorithm for calculating
$\sigma_z$ is given in \cite{matsakis97}.

Figure 4 plots the fractional instability of the intrinsic rotation of
the pulsars PSR B1855+09 and PSR B1937+21 taking into account the
contribution of the reference time and the variations TT--PT$_{\rm
  ens}$.  Theoretical curves for the behavior of  $\sigma_z$ when
noise due to the presence of a stochastic gravitational-wave
background with relative energy densities $\Omega_gh^2=10^{-9}$ and
$10^{-10}$ \cite{kaspi94} are also shown in the lowerright corner of
the figure. The  $\sigma_z$ curve crosses the line
$\Omega_gh^2=10^{-10}$; however, the upper limit for the
gravitational-wave background should be an order of magnitude higher,
taking into account the uncertainty in this quantity. Thus, as a
conservative estimate of this upper limit, we adopt $\Omega_gh^2
\lesssim 10^{-9}$.

\begin{figure}\label{fig4}
\setcaptionmargin{5mm}
\onelinecaptionstrue
\includegraphics[width=12cm]{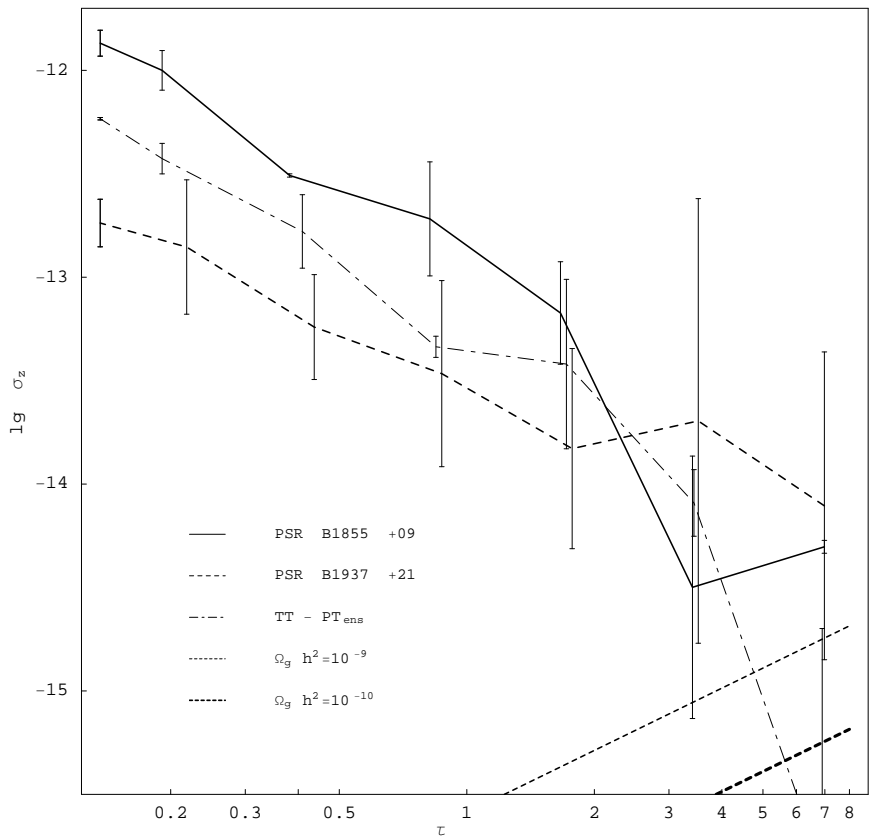} 
\captionstyle{normal}
\caption{Fractional instability $\sigma_z$, based on the chronometry
  data for PSR B1855+09 (solid) and PSR B1937+21 (dashed) and
  corresponding to TT--PT$_{\rm ens}$ (dot-dashed) as a
  function of the observing interval $\tau$ (in years). The
  theoretical curves of $\sigma_z$ for $\Omega_g h^2=10^{-9}$ and
  $10^{-10}$ are shown in the lower-right corner of the figure.}
\end{figure}

The fractional instability of TT(BIPM06) relative to PT$_{\rm ens}$ is at the
level $< 10^{-15}$ for a seven-year interval, and is an order of
magnitude better than the fractional instability of the rotation of
PSR B1855+09 and PSR B1937+21, taking into account the contribution of
the reference time system. As the numerical simulations show, when
Wiener filtration is used, the accuracy with which TT--PT$_{\rm ens}$ is
estimated grows with the number of pulsars used. We estimate the
accuracy of the optimal filtration method to be 170 ns.  This accuracy
was calculated as the rms deviation between the obtained and smoothed
signals. The smoothing was carried out in the frequency domain using a
low-frequency Kaiser filter with a bandwidth of $f_{\rm max}/32$, where
$f_{\rm max}=2/\Delta t$ and $\Delta t=10$ days is the sample
interval. Our choice of this bandwidth was based on the desire to
obtain smoothness similar to the UTC--TT(BIPM06) curve. Thus, the
uncertainty in our estimation of PT$_{\rm ens}$ can, in principle,
reach several tens of nanoseconds if the most stable millisecond
pulsars are used.

The proposed method does not distinguish quadratic trends in the
reference time and in the rotational phase of the pulsar, because the
pulsar itself displays secular deceleration of the rotational
frequency. In our opinion, this does not hinder the use of pulsars as
independent clocks, since longer-period variations in the reference
clocks will be revealed as data are accumulated over increasingly
longer time intervals.

The low relative accuracy of  $\dot P$ noted in \cite{guinot91}
likewise does not pose problems, since the rotational phase of the
pulsar is not predicted. However, if it is required to predict the
behavior of some reference atomic time, such as UTC, relative to the
group pulsar time, this can be done based on variations of
UTC--PT$_{\rm ens}$ using standard methods for the prognosis of time
series, provided that this prognosis is obtained relative to
short-period fluctuations, without linear and quadratic trends. In
this approach, the low relative accuracy of the derivative of the
rotational period of the pulsar will not play a role, since the
absolute phase is not predicted.

\section{CONCLUSION}

We have presented an algorithm for forming a group pulsar time based
on optimal Wiener filtration. The algorithm makes it possible to
distinguish the contributions to barycentric residual deviations of
PATs due to irregularities in the intrinsic rotation of the pulsar and
variations in the reference clocks.  Both irregularity in the pulsar
rotation and variations in the reference time scale are obtained
relative to an {\it ideal} time system. Realization of the algorithm
requires observations of at least two pulsars relative to a common
reference time.

The proposed approach has better accuracy than the use of weighted
averaging to form the group time scale, since it uses additional
information about the signal via its correlation function or power
spectrum.

The presence of a pulsar time that is independent of terrestrial
conditions makes it possible to carry out independent checks of
terrestrial time scales, which immediately provides possibilities for
revealing the presence of variations that are common to all
terrestrial standards that would otherwise be undetectable.

\begin{acknowledgments}
This work was partially supported by the joint Russian--Chinese project
''Study on X-ray Millisecond Pulsar Timing'' carried out at the
Pushchino Radio Astronomy Observatory, Astro Space Center, Lebedev
Physical Institute and the National Time Service Center of the Chinese
Academy of Sciences (NTSC), the Russian Foundation for Basic Research
(project 09-02-000584-a), and the Basic Research Program of the
Presidium of the Russian Academy of Sciences ''The Origin, Structure,
and Evolution of Objects in the Universe''.

\end{acknowledgments}

\appendix

\vspace{1cm}
Translated by D. Gabuzda
\end{document}